\documentclass[runningheads]{llncs}
\input{psfig.sty}
\begin{document}
%
%
\def\bsig{\mbox{\boldmath $\sigma$}}                          
\def\bsig{\mbox{\boldmath $\Sigma$}}
\def\bgam{\mbox{\boldmath $\gamma$}}
\def\bgam{\mbox{\boldmath $\Gamma$}}
\def\bphi{\mbox{\boldmath $\phi$}}
\def\bphi{\mbox{\boldmath $\Phi$}}
\def\btau{\mbox{\boldmath $\tau$}}
\def\btau{\mbox{\boldmath $\Tau$}}
\def\btau{\mbox{\boldmath $\partial$}}
\def\Delc{{\Delta}_{\circ}}
\def\bp{\mid {\bf p} \mid}
\def\al{\alpha}
\def\bet{\beta}
\def\gam{\gamma}
\def\del{\delta}
\def\Del{\Delta}
\def\te{\theta}
\def\nua{{\nu}_{\alpha}}
\def\nui{{\nu}_i}
\def\nuj{{\nu}_j}
\def\nue{{\nu}_e}
\def\num{{\nu}_{\mu}}
\def\nut{{\nu}_{\tau}}
\def\2te{2{\theta}}
\def\chic#1{{\scriptscriptstyle #1}}
\def\chicl{{\chic L}}
\def\lam{\lambda}
\def\SU{SU(2)_{\chic L} \otimes U(1)_{\chic Y}}
\def\Lam{\Lambda}
\def\sig{\sigma}
\def\'#1{\ifx#1i\accent19\i\else\accent19#1\fi}
\def\O{\Omega}
\def\o{\omega}
\def\s{\sigma}
\def\D{\Delta}
\def\d{\delta}
\def\df{\rm d}
\def\8{\infty}
\def\ld{\lambda}
\def\eps{\epsilon}
\def\ref#1{$^{#1}$}
\def\chicl{{\chic L}}
\def\lam{\lambda}
\def\SU{SU(2)_{\chic L} \otimes U(1)_{\chic Y}}
\def\Lam{\Lambda}
\def\sig{\sigma}
\def\'#1{\ifx#1i\accent19\i\else\accent19#1\fi}
\def\O{\Omega}
\def\o{\omega}
\def\s{\sigma}
\def\D{\Delta}
\def\d{\delta}
\def\df{\rm d}
\def\8{\infty}
\def\ld{\lambda}
\def\eps{\epsilon}
\def\ref#1{$^{#1}$}
\newcommand{\be}{\begin{equation}}
\newcommand{\ee}{\end{equation}}
\newcommand{\ba}{\begin{array}}
\newcommand{\ea}{\end{array}}
\newcommand{\dis}{\displaystyle}
\newcommand{\alfad}{\frac{\dis \bar \alpha_s}{\dis \pi}}
\mainmatter
\title{Appell Functions and the Scalar One-Loop Three-point 
Integrals in Feynman Diagrams}
\titlerunning{Appell Functions and the Scalar Three-point Functions}
\author{Luis G. Cabral- Rosetti\inst{1} \and Miguel A. Sanchis-Lozano\inst{2}}
\authorrunning{Luis G. Cabral- Rosetti and Miguel A. Sanchis-Lozano}
\institute{Instituto de Ciencias Nucleares, Departamento de F{\'\i}sica de 
Altas Energ{\'\i}as,\\
Universidad Nacional Aut\'onoma de M\'exico, (ICN-UNAM). Circuito Exterior, 
C.U., Apartado Postal 70-543, 94510 M\'exico, D.F., (M\'exico).\\
\email{luis@nuclecu.unam.mx}\\
\and
Departamento de F\'{\i}sica Te\'orica and IFIC, Centro Mixto\\
Universidad de Valencia-CSIC, 46100 Burjassot, Valencia (Spain)\\
\email{Miguel.Angel.Sanchis@ific.uv.es}}
\maketitle
\begin{abstract}
The scalar three-point function appearing in one-loop Feynman diagrams 
is compactly expressed in terms of a generalized hypergeometric function 
of two variables. Use is made of the connection between such Appell 
function and dilogarithms coming from a previous investigation. Special 
cases are obtained for particular values of internal masses and external 
momenta \footnote{Contribution to Proceeedings of the Second International 
Workshop on {\it Graphs, Operads, Logic, Parallel Computational and 
Mathematical Physics}, Cuautitl\'an, M\'exico, May 6 - May 16, 2002. This
paper is based on the Refs. \cite{mas1}, \cite{appell}.}.
\end{abstract}

\section{Introduction}

Recent and forthcoming unprecedented high-precision experimental results from 
$e^+e^-$ colliders (LEP, B Factories), hadron-hadron colliders (Tevatron, LHC)
and electron-proton colliders (HERA) are demanding refined calculations from 
the theoretical side as stringent tests of the Standard Model for the 
fundamental constituents and interactions in Nature. Moreover, possible 
extensions beyond the Standard Model ({\em e.g.} Supersymmetry) and
new physics ({\em e.g.} Extra dimensions) often require 
high-order calculations where such new effects eventually would manifest.

In fact, much effort has been devoted so far to develop systematic approaches 
to the evaluation of complicated Feynman diagrams, looking for algorithms 
(see for example \cite{hahn,brucher} and references therein), or recurrence 
algorithms \cite{mertig,tarasov2,tarasov1}, to be implemented in computer 
program packages to cope with the complexity of the calculation. On the other 
hand, the fact that such algorithms could be based, at least in part, on 
already defined functions represents a great advantage in many respects, as 
for example the knowledge of their analytic properties ({\em e.g.} branching 
points and cuts with physical significance), reduction to simpler cases with 
the subsequent capability of cross-checks and so on. Actually, commonly-used 
representations for loop integrals involve special functions like 
polylogarithms, generalized Clausen's functions, etc.

Furthermore, over this decade the role of generalized hypergeometric functions
in several variables to express the result of multi-leg and multi-loop
integrals arising in Feynman diagrams has been widely recognized 
\cite{davy0,davy,davy2,davy3,davy4,davy5,Schmidt,anastasiou,berends,berends2}.
The special interest in using hypergeometric-type functions is twofold:

\begin{itemize}
\item Hypergeometric series are convergent within certain domains
      of their arguments, physically related to some kinematic regions. This 
      affords numerical calculations, in particular implementations as 
      algorithms in computer programs. Moreover, analytic continuation allows 
      to express hypergeometric series as functions outside those convergence 
      domains.
\item The possibility of describing final results by means of known functions 
      instead of {\em ad hoc} power series is interesting by itself. This is 
      especially significant for hypergeometric functions because of their 
      deep connection with special functions whose properties are well 
      established in the mathematical literature. 
\end{itemize}

In this paper we mainly focus on the second point, although our aim is much 
more modest than searching for any master formula or method regarding complex 
Feynman diagrams. Rather we re-examine the scalar three-point function for 
massive external and internal lines, already solved in terms of dilogarithms 
\cite{olden}. The novelty of this work consists of solving and expressing the 
three-point function $C_0$, in terms of a set of Appell functions \cite{kampe}
whose arguments are combinations of kinematic quantities. Then, by using a 
theorem already proved in an earlier publication \cite{mas1}, we end up with 
sixteen dilogarithms, recovering a well-known result \cite{olden}. The elegant 
relationship shown in \cite{mas1} between Gauss and dilogarithmic functions 
might be a hint to look for more general connections that, hopefully, could be
helpful for finding compact expressions or algorithms in Feynman calculations.

\section{A Simple Connection Between Appel Functions and Dilogarithms}

Let us first write the main result published in \cite{mas1} which shows the 
relationship between dilogarithms and a special kind of Appell \vspace{0.1in}
function:

{\sc \underline {Theorem:}}

\be
\begin{array}{c}
\displaystyle
\frac{1}{2}xy\ F_{3}[1,1,1,1;3;x,y] = 
Li_{2}(x) + Li_{2}(y) - Li_{2}(x + y - xy)\ ,
\\[0.5cm]
\displaystyle
{\mid}\arg{(1-x)}{\mid}<\pi, \vspace{0.1in} 
{\mid}\arg{(1-y)}{\mid}<\pi\ 
and\ {\mid}\arg{(1-x)(1-y)}{\mid}<\pi.
\end{array}
\ee
\newline

{\sc \underline {Proof:}}

This formula can be proved by differentiating both sides with respect to $x$ 
and $y$ consecutively. Expanding the Appell's function \cite{kampe} as a 
double series, the result of differentiating the l.h.s. reads:

\be
\frac{1}{2}\ F_3[1,2,2,1;3;x,y]\ ;\ |x| < 1\ ,\ |y| < 1\ .
\ee

\noindent
Now, invoking the property \cite{exton}, \cite{lewin}, \cite{slater}:

\begin{center}
$F_3[\alpha,\gamma-\alpha,\beta,\gamma-\beta;\gamma;x,y]=
(1-y)^{\alpha+\beta-\gamma}\ _2F_1(\alpha,\beta;\gamma;x+y-xy)$
\end{center}

\noindent
which is valid in a suitable small open polydisc centered at the origin, 
we conclude that (2) can be rewritten as 
\[ \frac{1}{2}\ _2F_1(1,2;3;x+y-xy) \]

\noindent
Next, differentiating twice the r.h.s of Eq. (1) one gets:
\[ -\frac{1}{x+y-xy}\ \biggl[1+\frac{1}{x+y-xy}\ln{(1-(x+y-xy))}\biggr]\ =\ 
\ \frac{1}{2}\ _2F_1(1,2;3;x+y-xy) \]
the last step coming from the relation 
$z_2F_1(1,2;3;z)=-2[1+\frac{1}{z}\ln{(1-z)}]$. 
Then both sides in Eq. (1) would differ in $f(x)+g(y)$:
\[ \frac{1}{2}xy\ F_3[1,1,1,1;3;x,y]=Li_2(x)+Li_2(y)-Li_{2}(x+y-xy)+
f(x)+g(y) \] where $f(x)$ and $g(y)$ are functions to be determined by 
taking particular values of the variables. Setting $x=0$ (or $y=0$)
it is easy to see that $f(x)=g(y){\equiv}0$.

Now, by analytic continuation we dispense with the restriction on the small
polydisc, extending its validity to a suitable domain of $C^2$: in order to 
get a single-valued function, with a well-defined branch for each dilogarithm 
in Eq. (1), we assume further that 
${\mid}\arg{(1-x)}{\mid}<\pi$, ${\mid}\arg{(1-y)}{\mid}<\pi$ and
${\mid}\arg{(1-x)(1-y)}{\mid}<\pi$. (Observe that then each function of one 
complex variable obtained from (1) by fixing the other variable is analytic 
in the corresponding subset of $C^2$. Then the function of two variables 
$\frac{1}{2}xyF_3$ is analytic according to the theorem of Hartogs-Osgood.)

Let us note that the validity of expression (1) can be extended dropping
the restriction ${\mid}\arg{(1-x)(1-y)}{\mid}<\pi$ by slightly modifying
Eq. (1): 

\be
\begin{array}{c}
\displaystyle
\frac{1}{2}xy\ F_3[1,1,1,1;3;x,y]=Li_{2}(x)+Li_{2}(y)
\\[0.5cm]
\displaystyle
- Li_{2}(x+y-xy)-{\eta}(1-x,1-y)\ln{(x+y-xy)}
\end{array}
\ee

\noindent
where ${\mid}\arg{(1-x)}{\mid}<\pi$, ${\mid}\arg{(1-y)}{\mid}<\pi$ and 
${\eta}(a,b)=\ln{(ab)}-\ln{a}-\ln{b}$. The principal branches of the 
logarithm and dilogarithm are understood. As a particular but
illustrative case one finds

\begin{equation}
x\ F_3[1,1,1,1;3;x,y=1]\ =\ 2x\ _3F_2(1,1,1;2,2;x)\ =\ 2\ Li_2(x)
\end{equation}

\newpage

\begin{figure}[htb]
\centerline{
\psfig{figure=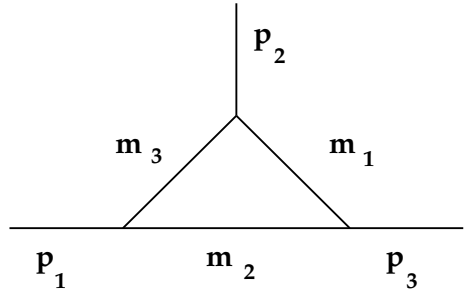,width=6.2cm}}
\caption{Notation used for external and internal momenta and masses}
\end{figure}

{\sc \underline {Corollary 1:}}

\begin{equation}
x^2\ F_3[1,1,1,1;3;x,-x]\ =\ Li_2(x^2)
\end{equation}
for $x$ \vspace{0.1in} real.\newline

{\sc \underline {Proof:}} 

It follows directly from Eq. (1) using the relation:

$$
Li_2(x)+Li_2(-x)=\frac{1}{2}Li_2(x^2)\ .
$$

{\sc \underline {Corollary 2:}}

\begin{equation}
x^2\ F_3[1,1,1,1;3;x,x] = 4 Li_2\biggl(\frac{1}{2-x}\biggr) + 
2\ln{}^2(2-x)\ -\ \frac{{\pi}^2}{3} 
\end{equation}
for $x$ real and less than unity.
\newline

{\sc \underline {Proof:}}

It follows directly from Eq. (1) using the relation:\\

$Li_2(2x-x^2)=2Li_2(x)-2Li_2 \Big( \frac{1}{2-x} \Big)
+ \frac{\pi^2}{6}-\ln{}^2(2-x)$ \ .\newline

{\sc \underline {Corollary 3:}}

\par
\begin{equation}
\lim_{y{\rightarrow}0}\ \frac{xy}{2}\ F_3[1,1,1,1;3;x,y]
= y \biggl[\ 1\ +\ \frac{1-x}{x}\ln{(1-x)} \biggr]
\end{equation}

{\sc \underline {Proof:}}

It follows directly from Eq. (1) using the relation: 
$ _2F_1(1,1;3;x) = (1-x)^{-1}$ $_2F_1(1,2;3;x/(x-1))$. 
If besides $x{\rightarrow}0$, the limit $xy/2$ is quickly recovered.

\begin{figure}
\centerline{\psfig{figure=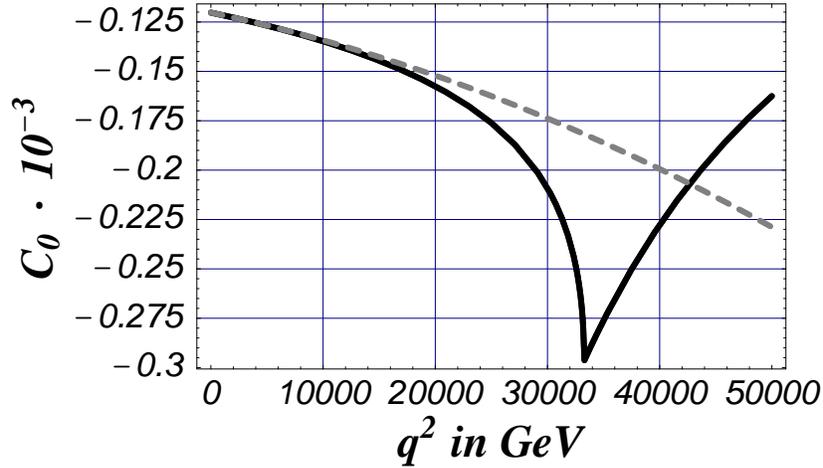,height=6.2cm}}
\caption{Comparation between exactly and aproximate solucions of
$C_0(q^2, 0, 0; 0, M_{\chic Z}^2, M_{\chic Z}^2)$. The solid
(black) line represent the exact solution Eq.~(12) valid for
{\it all $q^2$} and the dashed (grey) line represents the aproximate 
solution Eq.~(18) valid only when $q^2 \rightarrow 0$.}
\label{fig:example}
\end{figure}

\section{Evaluation of One-Loop Integrals}

Below we present an application of the relationship between
hypergeometric series and dilogarithms to the evaluation
of scalar three-point integrals appearing at one-loop level
in Feynman diagrams for field theoretical calculations.
\par
In particular we shall write the scalar three-point function
corresponding to the diagram of Figure 1:

\be
\begin{array}{c}
\displaystyle
C_0(p_1^2,p_2^2,p_3^3;m_1^2,m_2^2,m_3^2)\ =
\\[0.5cm]
\displaystyle
\frac{1}{(2\pi)^4}\ \int\ \frac{d^4q}{[q^2-m_2^2][(q+p_1)^2-m_3^3]
[(q+p_1+p_2)^2-m_1^2]}
\end{array}
\ee

\vspace{0.1in}
\noindent
After some manipulations \cite{mas2} the loop integral may be
written as

\be
\begin{array}{c}
\displaystyle
C_0(p_1^2,p_2^2,p_3^3;m_1^2,m_2^2,m_3^2) =  -\frac{i}{(4\pi)^2}
\frac{1}{\lambda^{1/2}(p_1^2,p_2^2,p_3^2)}\ {\times}
\\[0.5cm]
\displaystyle
\sum_{i=1}^{2}\ \biggl(\ [\ R(x_i,y)-R(x_i',y)\ ]\ -\ 
[\alpha{\leftrightarrow}(1-\alpha), 1{\leftrightarrow}3\ ]\ \biggr)
\end{array}
\ee

\noindent
where the notation is the \vspace{0.1in} following:

\begin{figure}
\centerline{\psfig{figure=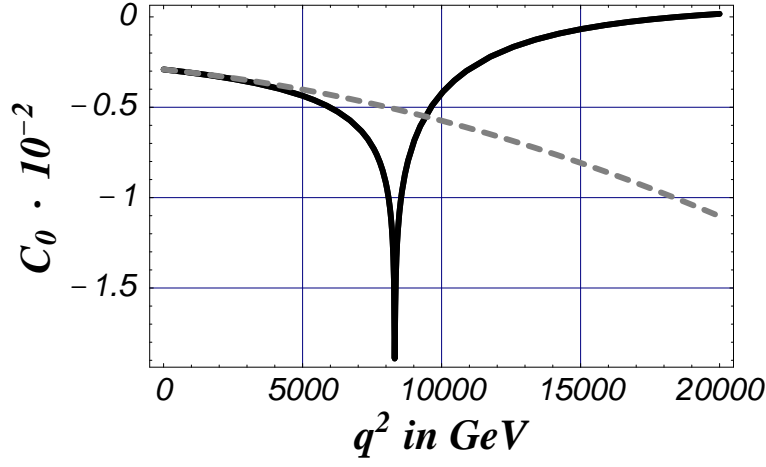,height=6.2cm}}

\caption{Comparation between exactly and aproximate solucions of
$C_0(q^2, 0, 0; m_e^2, M_{\chic Z}^2, 0)$. The solid
(black) line represents the exact solution Eq.~(14) valid for
{\it all $q^2$} and the dashed (grey) line represents the aproximate 
solution Eq.~(19) valid only when $q^2 \rightarrow 0$.}
\label{fig:example}
\end{figure}

\begin{itemize}
\item $R$ can be written in terms of an Appell's function according to
\[ R(x,y)\ =\ \frac{1}{2}\ xy\ F_3[1,1,1,1;3;x,y] \]
\item ${\lambda}$ stands for the K\"{a}llen function, 
$\lambda(x,y,z)= (x-y-z)^2-4yz$.
\item Assuming an (at least) external timelike momentum ${\alpha}$ is given by
any of the two solutions: \cite{thooft}
\[ \alpha_{\pm}\ =\ \frac{p_1^2+p_2^2-p_3^2{\pm}{\lambda}^{1/2}(p_1^2,p_2^2,p_3^2)}{2p_2^2} \]
\item Variables $x_i$ and $x_i'$ are defined as: $x_i=1/r_i$
and $x_i'={\alpha}/t_i$ where
\[ r_{1,2}\ =\ \frac{p_1^2+m_3^2-m_2^2{\pm}{\lambda}^{1/2}(p_1^2,m_3^2,m_2^2)}
{2p_1^2} \]
and \[ t_{1,2}\ =\ \frac{p_2^2+m_3^2-m_1^2{\pm}{\lambda}^{1/2}
(p_2^2,m_3^2,m_1^2)}{2p_2^2} \]
\item Variable $y$ is given by
\[ y\ =\ \frac{\alpha\ {\lambda}^{1/2}(p_1^2,p_2^2,p_3^2)}
{(1-\alpha)[-p_1^2+m_3^2-m_2^2]+\alpha[-p_3^2+m_1^2-m_2^2]} \]
\end{itemize}
\vskip 0.5 cm

Let us note the symmetry under the interchange of indices 
$1{\leftrightarrow}3$ and ${\alpha}_{\pm}{\leftrightarrow}(1-{\alpha}_{\mp})$ 
(where the subindices $\pm$ refer to the sign of the square root of above) 
implying the independence of the choice of any of both solutions for $\alpha$ 
as may be seen from (9), as a consequence of the choice of the external mass 
made zero in the procedure \cite{mas2}.

\noindent
Using the expression (1) the one-loop integral can be expressed in terms of 
sixteen dilogarithms since there is a pairlike cancellation between the eight 
$Li_2(y)$ functions. In Table~1, we present some special but relevant cases 
for the scalar three-point function ($C_0$).

\setlength{\tabcolsep}{4pt}
\begin{table}
\begin{center}
\caption{Special finite cases for the $C_0/i{\pi}^2$ three-point 
function \cite{appell}.}
\label{table:headings}
\begin{tabular}{ll}
\hline\noalign{\smallskip}
\centerline{
{\bf SOME SPECIAL CASES FOR THE THREE-POINT FUNCTION $C_0$}}\\
\noalign{\smallskip}
\hline
\noalign{\smallskip}
$C_0(0, 0, 0; m_{\chic 1}, m_{\chic 2}, m_{\chic 3}) = 
- \Big\{ \frac{m_{\chic 1}^2}{(m_{\chic 1}^2 - m_{\chic 2}^2)
(m_{\chic 1}^2 - m_{\chic 3}^2)} 
\ln \Big(\frac{m_{\chic 1}^2}{\mu^2} \Big)$\\ 
\ \ \ \ \ \ \ \ \ \ \ \ \ \ \ \ \ \ \ \ \ \ \ \ \ \ \ \ \ \ \ \ \ \ 
$+\ \frac{m_{\chic 2}^2}{(m_{\chic 2}^2 - m_{\chic 3}^2)
(m_{\chic 2}^2 - m_{\chic 1}^2)} 
\ln \Big(\frac{m_{\chic 2}^2}{\mu^2} \Big)$\\
\ \ \ \ \ \ \ \ \ \ \ \ \ \ \ \ \ \ \ \ \ \ \ \ \ \ \ \ \ \ \ \ \ \ 
$+\ \frac{m_{\chic 3}^2}{(m_{\chic 3}^2 - m_{\chic 1}^2)
(m_{\chic 3}^2 - m_{\chic 2}^2)} 
\ln \Big(\frac{m_{\chic 3}^2}{\mu^2} \Big) \Big\}$
\ \ \ \ \ \ \ \ \ \ \ \ \ \ \ \ \ \ \ \ \ \ \ \ \ \ \ \ \ \ \ (10)\\ \\
$C_0(p^2, 0, 0; m^2, 0, 0) = - \frac{1}{p^2} 
\Big\{ Li_{\chic 2} \Big( 1 + \frac{p^2}{m^2}\Big) 
- \zeta (2) \Big\}$
\ \ \ \ \ \ \ \ \ \ \ \ \ \ \ \ \ \ \ \ \ \ \ \ \ \ \ \ \ \ \ \ \ \ 
\ \ \ \ (11)\\ \\
$C_0(p^2, 0, 0; 0, m^2, m^2) = - \frac{2}{p^2}
\Big\{ Li_{\chic 2} \Big( \frac{1}{r_{\chic 1}} \Big) 
+\, Li_{\chic 2} \Big( \frac{1}{r_{\chic 1}} \Big) \Big\}$
\ \ \ \ \ \ \ \ \ \ \ \ \ \ \ \ \ \ \ \ \ \ \ \ \ \ \ \ \ \ \ \ \ \ 
\ (12)\\ \\
$C_0(p^2, 0, 0; m^2, m^2, m^2) = - \frac{1}{p^2}
\Big\{ Li_{\chic 2} \Big( \frac{1}{r_{\chic 1}} \Big) 
+\, Li_{\chic 2} \Big( \frac{1}{r_{\chic 1}} \Big) \Big\}$
\ \ \ \ \ \ \ \ \ \ \ \ \ \ \ \ \ \ \ \ \ \ \ \ \ \ \ \ \ \ \ \ (13)\\ \\
$C_0(p^2, 0, 0; m^2, M^2, 0) = - \frac{1}{p^2}
\Big\{ Li_{\chic 2} \Big( \frac{p^2 + m^2 - M^2}{m^2} \Big)
- Li_{\chic 2} \Big( \frac{m^2 - M^2}{m^2} \Big) \Big\}$
\ \ \ \ \ \ \ \ \ \ \ \ \ \ \ \ \ (14)\\ \\
$C_0(p^2, 0, 0; m^2, m^2, 0) = - \frac{1}{p^2}
Li_{\chic 2} \Big( \frac{p^2}{m^2} \Big)$
\ \ \ \ \ \ \ \ \ \ \ \ \ \ \ \ \ \ \ \ \ \ \ \ \ \ \ \ \ \ \ \ \ \ 
\ \ \ \ \ \ \ \ \ \ \ \ \ \ \ \ \ \ \ (15)\\ \\
$C_0(p_{\chic 1}^2, 0, p_{\chic 3}^2; m^2, 0, m^2) =
\frac{1}{p_{\chic 1}^2 - p_{\chic 3}^2} 
\Big\{ Li_{\chic 2} \Big( \frac{p_{\chic 3}^2}{m^2} \Big)
- Li_{\chic 2} \Big( \frac{p_{\chic 1}^2}{m^2} \Big) \Big\}$
\ \ \ \ \ \ \ \ \ \ \ \ \ \ \ \ \ \ \ \ \ \ \ \ \ \ \ \ \ (16)\\ \\
$C_0(p^2, 0, p^2; m^2, 0, m^2) = \frac{1}{p^2}
\ln \Big( 1 - \frac{p^2}{m^2} \Big)$
\ \ \ \ \ \ \ \ \ \ \ \ \ \ \ \ \ \ \ \ \ \ \ \ \ \ \ \ \ \ \ \ \ \ 
\ \ \ \ \ \ \ \ \ \ \ \ \ \ \ (17)\\ \\
\newline \\
where $\mu^2$ is the 't Hooft scale mass and 
$r_{\chic {1,2}} = \frac{1}{2} \Big[ 1 \pm 
\sqrt{1 - \frac{4 m^2}{p^2}}\ \Big]$\\
\newline \\
\hline
\end{tabular}
\end{center}
\end{table}
\setlength{\tabcolsep}{1.4pt}

\subsection{An Example of Numerical Evaluation}

Finally, we compare our exactly analytical solutions with the numerical 
evaluation of the scalar three point functions $C_0$ using the computer
program package {\it LoopTools} \cite{looptools}, as well as the aproximate
analytical expressions found with the {\it Taylor power series} near 
$q^2 = 0$, used in some phenomenological applications in neutrino physics
\cite{luis1}, \cite{luis2}.

In this regard, we have analyzed the Passarino-Veltman scalar 
three-point functions $C_0(q^2, 0, 0; 0, M_{\chic Z}^2, M_{\chic Z}^2)$ and 
$C_0(q^2, 0, 0; m_{\chic e}^2, M_{\chic Z}^2, 0)$ respectively, where
$q^2$ denotes the photon momentum, $m_{\chic e}$ and $M_{\chic Z}$ are the
electron and the vectorial boson $Z$ masses. The corresponding
plots can be seen in Figures 2 and 3. The solid (black) line, 
in both graphics, shows the exact solution of Equations (12) and (14),
perfectly matching the numerical evaluation using the aforementioned
{\it LoopTools} package. The dashed (grey) line represents the aproximate
solution of the {\it Taylor} expansion around $q^2 = 0$ whose expressions, 
up to ${\cal O}(q^6)$ order, are

$$
C_0(q^2, 0, 0; 0, M_{\chic Z}^2, M_{\chic Z}^2) \simeq
- \frac{1}{M_{\chic Z}^2} - \frac{q^2}{12\, M_{\chic Z}^4} 
- \frac{q^4}{90\, M_{\chic Z}^6} + {\cal O}(q^6)
\eqno{(18)}
$$

$$
\begin{array}{c}
\displaystyle
C_0(q^2, 0, 0; m_{\chic e}^2, M_{\chic Z}^2, 0) \simeq
\frac{\ln \Big( \frac{m_{\chic e}^2}{M_{\chic Z}^2} \Big)}
{M_{\chic Z}^2 - m_{\chic e}^2} 
+ \frac{q^2}{2\, (m_{\chic e}^2 - M_{\chic Z}^2)^3} 
\Big\{ (m_{\chic e}^2 - M_{\chic Z}^2) 
\Big[2 - \ln \Big( \frac{m_{\chic e}^2}{M_{\chic Z}^2} \Big) \Big] \Big\}
\\[0.5cm]
\displaystyle
+ \frac{q^4}{3\, (m_{\chic e}^2 - M_{\chic Z}^2)^5} 
\Big\{ 3\, (m_{\chic e}^4 - M_{\chic Z}^4) 
- (m_{\chic e}^4 + 4\, m_{\chic e}^2\, M_{\chic Z}^2 + M_{\chic Z}^4)\, 
\ln \Big( \frac{m_{\chic e}^2}{M_{\chic Z}^2} \Big) \Big\} + {\cal O}(q^6)
\end{array}
\eqno{(19)}
$$
\noindent
Note that our analytical solutions are valid for {\it all $q^2$}, in
contrast to the aproximate solution, only valid when $q^2 \rightarrow  0$,
which physically means $q^2<<M_Z^2$

\section{Summary and last remarks}
In this paper, we have shown that the familiar scalar three-point function 
$C_0$ arising in Feynman diagram calculations, can be expressed in terms
of eight Appell functions whose arguments are simple combinations of internal 
and external masses \cite{appell}. The extension to complex values of the 
masses (of concern  in several interesting physical cases) can be performed 
via analytic continuation of the hypergeometric functions. Moreover, by 
invoking a theorem proved in \cite{mas1} connecting dilogarithms and 
generalized hypergeometric functions, we have recovered a well-known formula 
for the evaluation of $C_0$ \cite{olden}, \cite{passa}. Finally, let us 
stress that this work is on the line of searching 
for closed expressions of Feynman loop integrals. On the other hand but in 
a complementary way, such physical requirements should motivate, on the 
mathematical side, further developments of the still largely unknown field 
on multiple hypergeometric functions (as compared to one-variable 
hypergeometric functions) and their connections to generalized 
polylogarithms \cite{kolbig}, \cite{tarasov3}.  

\subsubsection*{Acknowledgments}
M.A.S.L. has been partially supported by {\bf CICYT} (Spain) under grant 
AEN-99/0692 and L.G.C.R. has been supported in part by the grants: 
{\it Programa de Apoyo a Proyectos de Investigaci\'on e Innovaci\'on 
Tecnol\'ogica} ({\bf PAPIIT}) of the {\bf DGAPA-UNAM} {\it No. of Project}: 
{\sc IN109001} (M\'exico) and in part by the {\it Proyecto de Instalaci\'on} 
of the {\bf CoNaCyT} {\it No. of Project}: {\sc I37307-E} (M\'exico).

\end{document}